\documentclass[aps,prx,onecolumn,superscriptaddress,notitlepage,nopacs,amsmath,amstex,amssymb,citeautoscript,longbibliography,floatfix,letter,twocolumn]{revtex4-2}

\usepackage{enumitem}
\usepackage{natbib}
\usepackage[english]{babel}
\usepackage{letltxmacro}
\usepackage{latexsym}
\LetLtxMacro{\ORIGselectlanguage}{\selectlanguage}
\makeatletter
\DeclareRobustCommand{\selectlanguage}[1]{\@ifundefined{alias@\string#1}
    {\ORIGselectlanguage{#1}}
    {\begingroup\edef\x{\endgroup
       \noexpand\ORIGselectlanguage{\@nameuse{alias@#1}}}\x}}
\newcommand{\definelanguagealias}[2]{\@namedef{alias@#1}{#2}}
\makeatother
\definelanguagealias{en}{english}
\definelanguagealias{English}{english}

\usepackage{amsthm}

\usepackage{graphicx}
\usepackage{bm}
\usepackage{physics}
\usepackage[dvipsnames]{xcolor}
\usepackage{algorithm}
\usepackage{algpseudocode}
\usepackage{blindtext}
\usepackage[normalem]{ulem}
\makeatletter
\newcommand{\printfnsymbol}[1]{\textsuperscript{\@fnsymbol{#1}}}
\makeatother
\usepackage{hyperref}
\hypersetup{
    unicode=false,
    pdftoolbar=false,
    pdfmenubar=true,
    pdffitwindow=false,
    pdfstartview={FitH},
    pdftitle={},
    pdfauthor={},
    pdfsubject={},
    pdfcreator={},
    pdfproducer={},
    pdfkeywords={},
    pdfnewwindow=true,
    colorlinks=true,
    linkcolor=teal,
    citecolor=teal,
    filecolor=teal,
    urlcolor=teal
}

\begin{document}
\title{Integrability breaking and bound states in Google's decorated XXZ circuits}

\author{Ana Hudomal}
\affiliation{School of Physics and Astronomy, University of Leeds, Leeds LS2 9JT, United Kingdom}
 \affiliation{Institute of Physics Belgrade, University of Belgrade, 11080 Belgrade, Serbia}

\author{Ryan Smith}
\affiliation{School of Physics and Astronomy, University of Leeds, Leeds LS2 9JT, United Kingdom}

\author{Andrew Hallam}
\affiliation{School of Physics and Astronomy, University of Leeds, Leeds LS2 9JT, United Kingdom}

\author{Zlatko Papi\'c}
\affiliation{School of Physics and Astronomy, University of Leeds, Leeds LS2 9JT, United Kingdom}

\date{\today}
\begin{abstract} 
Recent quantum simulation by Google [Nature {\bf 612}, 240 (2022)] has demonstrated the formation of bound states of interacting photons in a quantum-circuit version of the XXZ spin chain. While such bound states are protected by integrability in a one-dimensional chain, the experiment found the bound states to be unexpectedly robust when integrability was broken by decorating the circuit with additional qubits, at least for small numbers of qubits ($\leq 24$) within the experimental capability. Here we scrutinize this result by state-of-the-art classical simulations, which greatly exceed the experimental system sizes and provide a benchmark for future studies in larger circuits. We find that the bound states consisting of a small and \emph{finite} number of photons are indeed robust in the non-integrable regime, even after scaling to the infinite time and infinite system size limit. Moreover, we show that such systems possess unusual spectral properties, with level statistics that deviates from the random matrix theory expectation. On the other hand, for low but finite density of photons, we find a much faster onset of thermalization and significantly weaker signatures of bound states, suggesting that anomalous dynamics may only be a property of dilute systems with \emph{zero} density of photons in the thermodynamic limit. 
The robustness of the bound states is also influenced by the number of decoration qubits and, to a lesser degree, by the regularity of their spatial arrangement.
\end{abstract}
\maketitle


\section{Introduction}

Recent advances in quantum simulators based on  ultracold atoms, trapped ions and superconducting circuits~\cite{Bloch2008,Bloch2012,Georgescu2014,Kjaergaard,Blatt2012,Houck2012,MonroeRMP,Browaeys2020} have opened a window to studying far-from-equilibrium dynamics and thermalization in isolated many-body systems~\cite{PolkovnikovRMP,Gogolin2016,dAlessio2016,Ueda2020}. 
The behavior of generic thermalizing systems is described by the Eigenstate Thermalization Hypothesis (ETH)~\cite{DeutschETH, SrednickiETH, RigolNature} which seeks to explain the process of thermalization at the level of the system's energy eigenstates.  In certain systems, the ETH can break down, allowing for new types of dynamical behavior and phases of matter to emerge~\cite{Huse-rev}.  One of the most striking manifestations of ergodicity breakdown occurs in finely tuned one-dimensional systems~\cite{Kinoshita06}, which fail to thermalize due to their rich symmetry structure known as quantum integrability~\cite{Sutherland,Bertini2021}. 

A paradigmatic quantum-integrable system is the spin-1/2 XXZ model, which describes the low-energy physics of certain ferromagnetic materials~\cite{MattisBook}. In one spatial dimension, the model's \emph{tour de force} analytic solution in the isotropic limit was presented by Bethe in the 1930s~\cite{Bethe1931}. One remarkable consequence of that solution was a special class of eigenstates that can be viewed as bound states of magnons -- the elementary quasiparticle excitations, whose signatures were observed in spectroscopic experiments~\cite{Date1966,Torrance1969,Hoogerbeets1984}. 
However, due to the challenges of probing bound states via conventional techniques such as inelastic neutron scattering, it has been proposed~\cite{Ganahl2012} that local quenches~\cite{Gobert2005,Petrosyan2007,Langer2009,Ren2010,Santos2011,Steinigeweg2011,Langer2011} may provide deeper insight into the physics of bound states~\cite{Pereira2008,Pereira2009,Caux2005a,Caux2005b,Kohno2009,Shashi2011,Imambekov2012}. Dynamical signatures of bound states were indeed observed in systems of $\mathrm{^{87}Rb}$ atoms in an optical lattice, realizing an effective Heisenberg model~\cite{Fukuhara2013}. 

While previous studies mostly focused on systems with continuous dynamics governed by a static Hamiltonian, it is also possible to construct equivalent Floquet models defined as a product of unitary matrices. 
Such models, whose quantum dynamics is intrinsically discrete, are better suited for quantum simulators which operate as a sequence of unitary gates.
Quantum circuit models that correspond to the spin-1/2 Heisenberg model in the high frequency limit were studied in Refs.~\cite{Vanicat2018,Ljubotina2019}.
Remarkably, the Floquet circuit realization was shown to be integrable for arbitrary parameters and not only in the small time step limit where it reduces via Troterrization to the Hamiltonian model~\cite{Vanicat2018,Ljubotina2019,Claeys2022}. 

The Floquet XXZ model was recently experimentally realized using a ring of superconducting qubits connected by high-fidelity fSim quantum logic gates~\cite{Google}.
These qubits interact with each other by superconducting currents and can host excitations in the form of trapped photons.
This setup has allowed for the preparation and observation of bound states of a few interacting photons,  which were predicted and analytically studied in Ref.~\cite{Aleiner2021}. One of the advantages was the possibility of controllably breaking the integrability by attaching extra qubits to the main chain and thus changing the geometry of the system.  In contrast with the expectation that the bound states are protected by integrability, it was experimentally observed that these states survive even in the non-integrable regime, as previously suggested for the Hamiltonian version of the model~\cite{Ganahl2012}. However, the robustness of the bound states was not studied in detail and the question of which mechanism protects it in the non-integrable case remains open. 

In this work, we use classical simulations, based on exact diagonalization (ED) and matrix product states (MPS), to gain understanding of the experiment in Ref.~\cite{Google}. Specifically, we study the statistical properties of the Floquet spectrum in order to detect the transition from integrable to the non-integrable regime. We also employ time-evolving block decimation (TEBD) simulations to investigate the evolution of bound states and their robustness. In this way, we are able to reach far larger system sizes, photon numbers, and timescales compared to the quantum hardware~\cite{Google}. In contrast to the experiment, which has limitations  due to the unwanted leakage of photons, the photon number is conserved in our study. We find that sectors with \emph{small but fixed} photon number have non-thermalizing spectral properties, which affect both their level statistics and quantum dynamics. Additionally, we confirm the  experimental finding that the bound states in these sectors persist beyond the integrable regime. While this effect is pronounced in dilute systems containing small photon numbers, it appears to be restricted to \emph{zero density} of excitations in the thermodynamic limit. By contrast, sectors with small but finite excitation density are found to thermalize rapidly as the photon number is increased, in parallel with the fast decay of bound states.

The remainder of this paper is organized as follows. In Sec.~\ref{sec:model} we introduce the Floquet XXZ model that will be the main object of our study. In Sec.~\ref{sec:integrability} we compute the corresponding Floquet modes and investigate the statistical properties of their energy levels, including the average ratio of consecutive energy gaps, the density of states and the spectral form factor.  
In Sec.~\ref{sec:bound_states} we study the evolution of bound states and their robustness to integrability breaking. We perform extrapolations to infinite system size and compare the data against the diagonal ensemble predictions, which provides information about the \emph{infinite-time} limit.
In Sec.~\ref{sec:other} we discuss several cases beyond those studied in experiment, in particular systems with a constant filling factor and different decoration patterns, including non-symmetric ones. We summarize our results and discuss their implications in Sec.~\ref{sec:discussion}. Appendices provide more details about the corresponding continuous XXZ model, effects of different parameters,
and the number of special eigenstates which affect the level statistics.

\section{Model}\label{sec:model}

\begin{figure}[tb]
 \includegraphics[width=\linewidth]{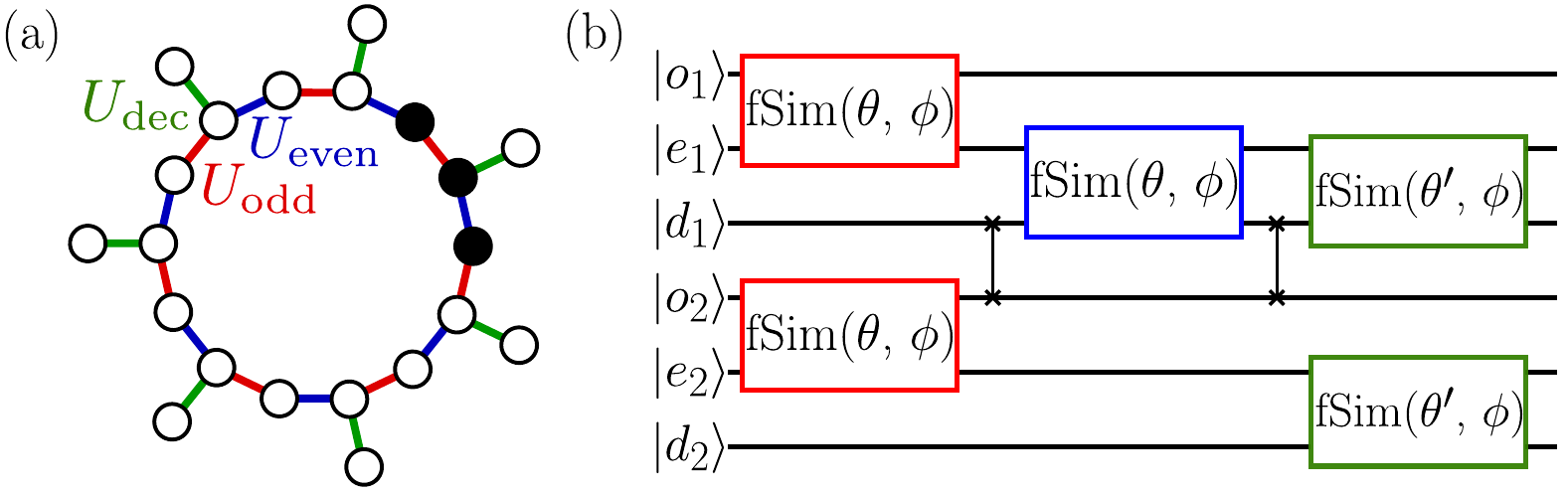}
 \caption{(a) Sketch of the XXZ circuit model with $L{=}14{+}7$ sites (7 unit cells). Filled dots denote a bound state of $N{=}3$ photons. Here, the integrability-breaking decorations are attached to every other site.
 (b) An example of the corresponding quantum circuit with $L{=}4{+}2$ sites. The circuit consists of fSim (boxes) and SWAP gates (vertical lines).  The alternating layers of gates acting on even/odd bonds and decorations are denoted by blue, red and green color, respectively, matching the unitaries $U_\mathrm{even}$, $U_\mathrm{odd}$ and $U_\mathrm{dec}$ in (a). Our classical MPS simulations in iTensor~\cite{itensor} follow this diagram and assume open boundary conditions. 
 }\label{fig:model}
 \end{figure}

The experiment from Ref.~\cite{Google} has realized a decorated ring of  superconducting qubits, schematically illustrated in  Fig.~\ref{fig:model}(a). 
If the occupancy is limited to zero or one photon per qubit, the photons can be modeled as hard-core bosons. Since we are considering a ring of qubits, we will impose periodic boundary conditions (PBCs) in our ED calculations, unless stated otherwise. The fundamental building block of the circuit is a 2-qubit fSim gate acting on pairs of adjacent qubits,
\begin{equation}\label{eq:fSim}
\mathrm{fSim}(\theta,\phi,\beta)=
\begin{pmatrix}
1 & 0 & 0 & 0\\
0 & \cos\theta & ie^{i\beta}\sin\theta & 0\\
0 & ie^{-i\beta}\sin\theta & \cos\theta & 0\\
0 & 0 & 0 & e^{i\phi}
\end{pmatrix},
\end{equation}
where $\theta$ and $\beta$ determine the nearest-neighbor hopping amplitude and phase, while $\phi$ represents the strength of interactions between neighboring qubits. The parameter $\beta$ mimics the external magnetic flux threading the ring.
In the following, we will primarily consider the case $\mathrm{fSim}(\theta,\phi,\beta{=}0)=\mathrm{fSim}(\theta,\phi)$. 

 Fig.~\ref{fig:model}(a) is a sketch of the model with decorations attached to every other site as in Ref.~\cite{Google}.  The number of photons will be denoted by $N$ and the total number of sites by $L{=}L_\mathrm{sites}{+}L_\mathrm{decor}$, which includes both the sites on the ring $L_\mathrm{sites}$ and the extra sites $L_\mathrm{decor}$.
 The sketch also depicts a state with $N{=}3$ adjacent photons, which will typically be used as the initial state in our simulations. Note that there is another, similar configuration of three adjacent photons, that is simply shifted by one lattice site. This configuration is inequivalent to the one in  Fig.~\ref{fig:model}(a) because it is connected to two decorations instead of one. As specified below, we will occasionally find it useful to average the results over these two initial states. In addition to the layout shown here, in Sec.~\ref{sec:other} we will also consider other decoration patterns. In general, we find the dynamical properties are highly sensitive to the number of photons and the decoration pattern.

Fig.~\ref{fig:model}(b) shows the corresponding quantum circuit which consists of fSim and SWAP gates. The states of the even, odd and decoration qubits are denoted by $\lvert {e}_i\rangle$, $\lvert {o}_i\rangle$ and $\lvert {d}_i\rangle$, respectively. Our classical TEBD simulations follow the layout in Fig.~\ref{fig:model}(b) and, for convenience, assume open boundary conditions (OBCs). We emphasize that the results below are insensitive to the choice of boundary conditions, as we will demonstrate good agreement between TEBD with OBCs and ED with PBCs.  
The circuit is defined by first applying fSim gates across all odd bonds, then across all even bonds. Since the even and odd bonds are thus not equivalent, the system is invariant to translation by two lattice sites of the main chain.
Additional gates which couple to the integrability-breaking extra sites $\lvert {d}_i\rangle$ are subsequently applied, which can further reduce the symmetry of the full system depending on the pattern of arrangement of the extra sites. 
The coupling parameter $\theta'$ is used to tune between the integrable and non-integrable regimes, while the interaction strength $\phi=\phi'$ is the same, both along the main chain and between the chain and the decorations.
The one-cycle unitary operator is then
\begin{align}\label{eq:Uf}
    \hat{U}_F=&\underbrace{\prod_\text{extra bonds}\mathrm{fSim}(\theta',\phi)}_{\hat{U}_\text{dec}}\times\nonumber\\
    &\underbrace{\prod_\text{even bonds}\mathrm{fSim}(\theta,\phi)}_{\hat{U}_\text{even}}
    \underbrace{\prod_\text{odd bonds}\mathrm{fSim}(\theta,\phi)}_{\hat{U}_\text{odd}}.
\end{align}
As shown in Appendix~\ref{appendix:XXZ}, the Hamiltonian of the XXZ model corresponds to the Trotter-Suzuki expansion of this unitary in the $\phi,\theta,\theta'\rightarrow0$ limit. 

The isotropic XXX version of the model in Eq.~(\ref{eq:Uf}) was first proposed in Ref.~\cite{Vanicat2018}, while the Floquet XXZ model was formulated in Ref.~\cite{Ljubotina2019} and analytically studied in detail in Ref.~\cite{Aleiner2021}.  The latter used Bethe ansatz to derive the dispersion of bound states containing an arbitrary number of photons.  These bound states are formed by stable magnon quasiparticles, and there are two different phases depending on the ratio of $\theta$ and $\phi$: (1) gapped phase $\phi>2\theta$, where bound states of any photon number exist for any momentum, and (2) gapless phase $\phi<2\theta$ where the bound states are only present for a finite range of momenta. The maximal group velocity was found to decrease with the number of photons in the bound state. Quantum simulations~\cite{Google} have later confirmed the analytical relations between the velocity of quasiparticles and their momentum.  
However, analytical solutions are not available for the non-integrable case, where the integrability is broken either by adding certain perturbations or by changing the geometry of the system. We will use classical simulations to numerically study this regime.

\subsection{Circular orthogonal ensemble}\label{subsec:symmetry}

 Before we analyze in detail the Floquet spectrum of Eq.~\eqref{eq:Uf}, we must understand the relevant symmetries of the model as they affect the random matrix theory ensemble describing the spectrum after breaking the integrability~\cite{Mehta2004}. For example, the undecorated model with PBCs is invariant under translation by two sites, due to the even and odd layers of fSim gates being applied separately. Such a circuit is also invariant under spatial inversion. However, attaching extra qubits to some of the sites reduces the symmetry of the full system. Regular patterns with decorations on every $n$th site will preserve some form of translation invariance, although with a larger unit cell. Furthermore, the system can be inversion-symmetric only if the decoration pattern itself is also inversion-symmetric. However, in some cases, such as that with decorations on every other site, the inversion of the decorations can be incompatible with the inversion of the main ring due to different reflection axes, so the full system only has translation symmetry, even though the decoration pattern is still inversion-symmetric. This will be discussed in more detail in Section~\ref{subsec:patterns}.
 
For a general unitary matrix $\hat{U}_F$, the level statistics is expected to conform to the Circular Unitary Ensemble (CUE). However, as will be apparent 
in Secs.~\ref{subsec:levelstat} and \ref{subsec:patterns}, in most cases studied here we obtain the Circular Orthogonal Ensemble (COE) statistics instead. COE would trivially ensue if $\hat{U}_F=\hat{U}_F^T$, however this is not the case here for any arrangement of decorations. Our calculations show that the necessary conditions for COE level statistics are an inversion-symmetric decoration pattern and equal parameters for the fSim gates on the even and odd bonds along the ring, as defined in Eq.~\eqref{eq:Uf}. 
 Additionally, the mirror axis for inversion needs to be centred on a site, not on a bond between two sites. If $\hat{R}$ is the inversion symmetry operator which reflects the qubits along this axis, we well then have $\hat{R}\hat{U}_\text{odd}\hat{R}=\hat{U}_\text{even}$, $\hat{R}\hat{U}_\text{even}\hat{R}=\hat{U}_\text{odd}$ and $\hat{R}\hat{U}_\text{dec}\hat{R}=\hat{U}_\text{dec}$. For simplicity, we define a modified one-cycle unitary operator
\begin{align}\label{eq:Uf2}
\hat{U}'_F=\sqrt{\hat{U}_\text{dec}}\hat{U}_\text{even}\hat{U}_\text{odd}\sqrt{\hat{U}_\text{dec}}.
\end{align}
The operators $\hat{U}_F$ and $\hat{U}^\prime_F$ have the same spectrum, since they differ only by a time shift.
It is now easy to see that $\hat{U}^\prime_F=\hat{R}\hat{U}^{\prime T}_F\hat{R}$.
This can be understood as an additional symmetry which relates the evolution operator and its transpose, resulting in COE level statistics.
Our situation is reminiscent of Ref.~\cite{Regnault2016}, where the Floquet spectrum was shown to have COE instead of CUE statistics if there is a transformation which connects the two steps of the Floquet unitary.

Another possibility is when the mirror axis is between two adjacent sites, leading to $\hat{R}\hat{U}_\text{odd}\hat{R}=\hat{U}_\text{odd}$ and $\hat{R}\hat{U}_\text{even}\hat{R}=\hat{U}_\text{even}$. We will then have $\hat{U}'_F=\hat{R}\hat{U}'_F\hat{R}$, meaning that $\hat{R}$ is simply another symmetry of $\hat{U}'_F$ which needs to be resolved. The level statistics in the sector where $\hat{R}$ has eigenvalue ${+}1$ is then CUE. We only find deviations from this expectation for small numbers of decorations such as two or four adjacent decorations, where the level statistics after resolving the $\hat{R}$ symmetry is somewhere between COE and CUE. However, it seems to increase towards CUE as the density of decorations or the number of photons is increased. 
There are also special cases which are inversion symmetric in respect to both types of mirror axes, such as the pattern with decorations on every third site. In those cases the level statistics stays COE even after resolving the $\hat{R}$ symmetry. In contrast, all non-symmetric decoration arrangements were found to exhibit CUE level statistics.

\section{Spectral properties}\label{sec:integrability}

In this section we analyze the spectrum of our unitary circuit model in Eq.~(\ref{eq:Uf}). This model does not have a Hamiltonian representation in the general case, since the mapping to the XXZ model (Appendix~\ref{appendix:XXZ}) is only valid in the $dt{\rightarrow}0$ limit. As a consequence, the system does not have eigenstates in the usual sense. However, we can instead compute the eigenstates of the one-cycle evolution operator $\hat{U}_F$~\eqref{eq:Uf}, which are known as the  Floquet modes. The corresponding Floquet quasienergy spectrum is periodic with periodicity $2\pi/T$, where $T$ is the time length of one cycle. We set the units such that $T=1$. We will investigate the properties of the Floquet modes and quasienergies from two complementary perspectives. On the one hand, we will study the level statistics and density of states, which directly derive from the quasienergies and thus tell us about the behavior of the system at very late time scales corresponding to the Heisenberg time. On the other hand, we will contrast these results against the spectral form factor, which provides information about intermediate time scales relevant for transport, such as the Thouless time. 

\subsection{Level statistics}\label{subsec:levelstat}
 
\begin{figure}[bth]
\includegraphics[width=0.49\textwidth]{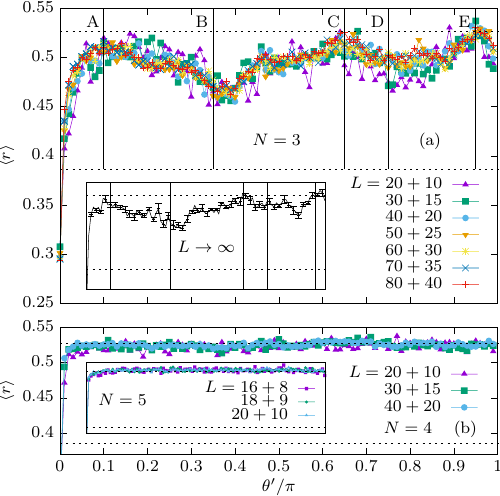} 
 \caption{Statistics of the Floquet quasienergies. Average ratio of consecutive energy gaps $\langle r\rangle$ for different values of $\theta'$ with fixed $\theta=\pi/6$, $\phi=2\pi/3$, $\beta=0$. The horizontal dashed lines are $\langle r\rangle_\mathrm{P}\approx0.386$ and $\langle r\rangle_\mathrm{COE}\approx0.527$.
 (a) $N=3$ photons for different system sizes indicated in the legend. The relevant Hilbert space dimensions range from   $\mathrm{dim}_{L{=}20{+}10}=406$ to $\mathrm{dim}_{L{=}80{+}40}=7021$. 
 The vertical lines A, B, C, D and E mark several values of $\theta'$ (0.10, 0.35, 0.65, 0.75 and 0.95) that will be studied later in more detail.
 Inset: linear extrapolation to $L{\rightarrow}\infty$.
 (b) The case of $N=4$ photons and (inset) $N=5$ photons, with the largest Hilbert space dimensions $\mathrm{dim}_{L{=}40{+}20}=24405$ and $\mathrm{dim}_{L{=}20{+}10}=14253$, respectively.
 In all the plots, we resolve the translation symmetry and consider only the $k{=}0$ momentum sector. 
 }\label{fig:level_statistics}
 \end{figure}

In order to determine whether our model Eq.~(\ref{eq:Uf}) is integrable or chaotic, we study the statistics of its quasienergy levels. In particular, we examine the level statistics ratio, $r=\min(s_n,s_{n+1})/\max(s_n,s_{n+1})$, characterizing the spacing of adjacent quasienergy gaps $s_n=\epsilon_{n+1}-\epsilon_{n}$~\cite{OganesyanHuse}. 
An integrable system is expected to follow the Poisson distribution with the average value $\langle r\rangle_\text{P} \approx 0.386$, while in the chaotic regime the expected distribution for our case, as explained in Sec.~\ref{subsec:symmetry} above, is the Circular Orthogonal Ensemble (COE) with $\langle r\rangle_\text{COE} \approx 0.527$
~\cite{Mehta2004,DAlessio2014}. We vary the hopping amplitude $\theta'$ between the main chain and the extra sites from $0$ to $\pi$ and plot the corresponding $\langle r\rangle(\theta')$. 
Fig.~\ref{fig:level_statistics}(a) shows the results for $N=3$ photons for various chain lengths, while the extrapolation to an infinitely large system $L{\rightarrow}\infty$ is plotted in the inset. This result should be contrasted against the results for $N=4$ and $N=5$ photons in Fig.~\ref{fig:level_statistics}(b).

Turning on the coupling to the decorations is expected to break integrability, which can indeed be observed in Fig.~\ref{fig:level_statistics} where the value of $\langle r\rangle(\theta')$ rapidly jumps towards $\langle r\rangle_\text{COE}$ as soon as $\theta'\neq0$. For $N\geq4$ photons, as soon as $\theta'{\gtrsim} 0.05\pi$, the level statistics becomes pinned to the COE value, in agreement with the usual expectation for integrability breaking in Hamiltonian systems~\cite{Santos2010}. 
However, the case with $N=3$ photons shows a visible departure from these expectations, exhibiting pronounced dips towards the Poisson value at special values of $\theta^\prime$ -- see Fig.~\ref{fig:level_statistics}(a). Furthermore, we find that the positions of the dips in $\langle r \rangle$ depend on the main chain hopping amplitude $\theta$, but not on the interaction strength $\phi$ or the flux through the ring $\beta$, see Appendix~\ref{appendix:parameters}. No emergent symmetry which would explain the dips at certain values of $\theta'$ could be identified. Instead, we will relate the presence of dips with special structures in the density of states in Sec.~\ref{subsec:dos} below.
 
We note that in all cases plotted in Fig.~\ref{fig:level_statistics}, the value of $\langle r\rangle(0)$ lies below the Poisson line, even though the model is known to be integrable at $\theta^\prime=0$. This is simply due to a large number of degeneracies present in the Floquet spectrum, which originate from the decorations and produce a peak in the probability distribution for zero level spacing. Even though there is no hopping to the extra sites when $\theta'=0$, there are still states where one or more photons are frozen in these additional sites. A state with all photons outside the main chain has zero energy, as do some states with two separate photons on the main chain and all other photons outside. We found that completely removing the extra sites brings $\langle r\rangle(0)$ closer to $\langle r\rangle_\text{P}$. 
We also note that, while the hopping amplitudes inside the main chain and between the chain and the extra sites are different, $\theta\neq\theta'$, the nearest-neighbor interaction strength is equal in both cases $\phi=\phi'$. This means that the photons frozen in decorations can still interact with the other photons. 

Finally, we emphasize that in Fig.~\ref{fig:level_statistics} we assumed a fixed decoration pattern chosen in Ref.~\cite{Google}, where an extra site is attached to every other main-chain site, such that the system is invariant to translations by two sites. In Sec.~\ref{subsec:patterns} we consider other decoration patterns, including the case of a single decoration, three decorations on the second, fourth and sixth site, and a random arrangement of decorations on half of all sites. 
All of these patterns break the translation symmetry and do not exhibit oscillations in $\langle r\rangle(\theta')$ that are visible in Fig.~\ref{fig:level_statistics}(a).
Instead, $\langle r\rangle$ first reaches a plateau 
and then starts to decay at larger values of $\theta'$.
The plateau is at $\langle r\rangle_\mathrm{COE}$ for inversion-symmetric patterns and at $\langle r\rangle_\mathrm{CUE}$ for non-symmetric ones.
We have also considered patterns which preserve some form of translation symmetry, for example those with decorations on every site, or every third, fourth or fifth site. The level statistics for these patterns shows similar properties to the previous case of decorations on every other site, with deviations from $\langle r\rangle_\mathrm{COE}$ for $N=3$, albeit with minima and maxima in $\langle r\rangle$ at different locations. In contrast, no such deviations were observed for $N\geq4$.

\subsection{Density of states}\label{subsec:dos}

\begin{figure}[bt]
\includegraphics[width=0.49\textwidth]{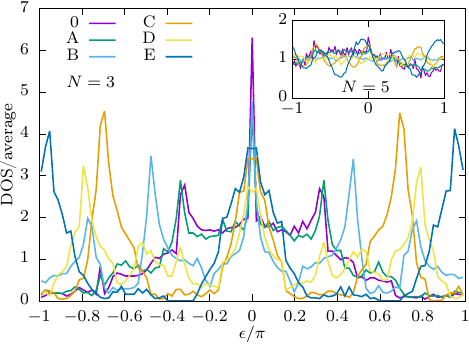}
 \caption{Density of states (DOS) for different values of $\theta'$, normalized by the average over the entire energy spectrum [the labels A--E are defined in Fig.~\ref{fig:level_statistics}(a), while 0 denotes the integrable case $\theta^\prime=0$]. The main panel corresponds to $N=3$ photons in a system size $L=80+40$, while the inset shows $N=5$ in $L=20+10$. In both cases,  $\theta=\pi/6$, $\phi=2\pi/3$. 
 }\label{fig:dos}
 \end{figure}

The intriguing features in the level statistics observed in Fig.~\ref{fig:level_statistics}(a) can be understood from the density of states (DOS). For example, sharp peaks in DOS signal a large number of degeneracies in the spectrum, which can decrease the value of $\langle r\rangle$. In Fig.~\ref{fig:dos}, we plot the normalized DOS curves for $N=3$ and $N=5$ photons at several values of $\theta'$ that were marked by A-E in Fig.~\ref{fig:level_statistics}(a). Both photon numbers exhibit a peak at $\epsilon{=}0$ when $\theta'{=}0$, which is explained by the previously discussed large number of zero modes due to the extra sites. This zero-energy peak is much more prominent for $N=3$ and its relative height decreases with $N$. The results for $N=4$ (not shown) are in between those for $N=3$ and $N=5$, with more peaks than $N=5$, but still overall flatter than $N=3$.
As $\theta'$ is increased, the DOS curves become more flat. 
However, several other notable peaks are present for $N=3$.
Although these peaks are visible at all $\theta'$, they are particularly sharp at those values where $\langle r\rangle$ deviates from $\langle r\rangle_\text{COE}$ (e.g. $\theta'\in[0.25\pi,0.45\pi]$ and $\theta'\in[0.75\pi,0.85\pi]$), see Figs.~\ref{fig:level_statistics}(a) and \ref{fig:dos}. The peaks in DOS are not present for non-symmetric patterns of extra sites, such as just one or three decorations, which will be discussed in Sec.~\ref{subsec:patterns}. 

\begin{figure}[bt]
 \includegraphics[width=0.49\textwidth]{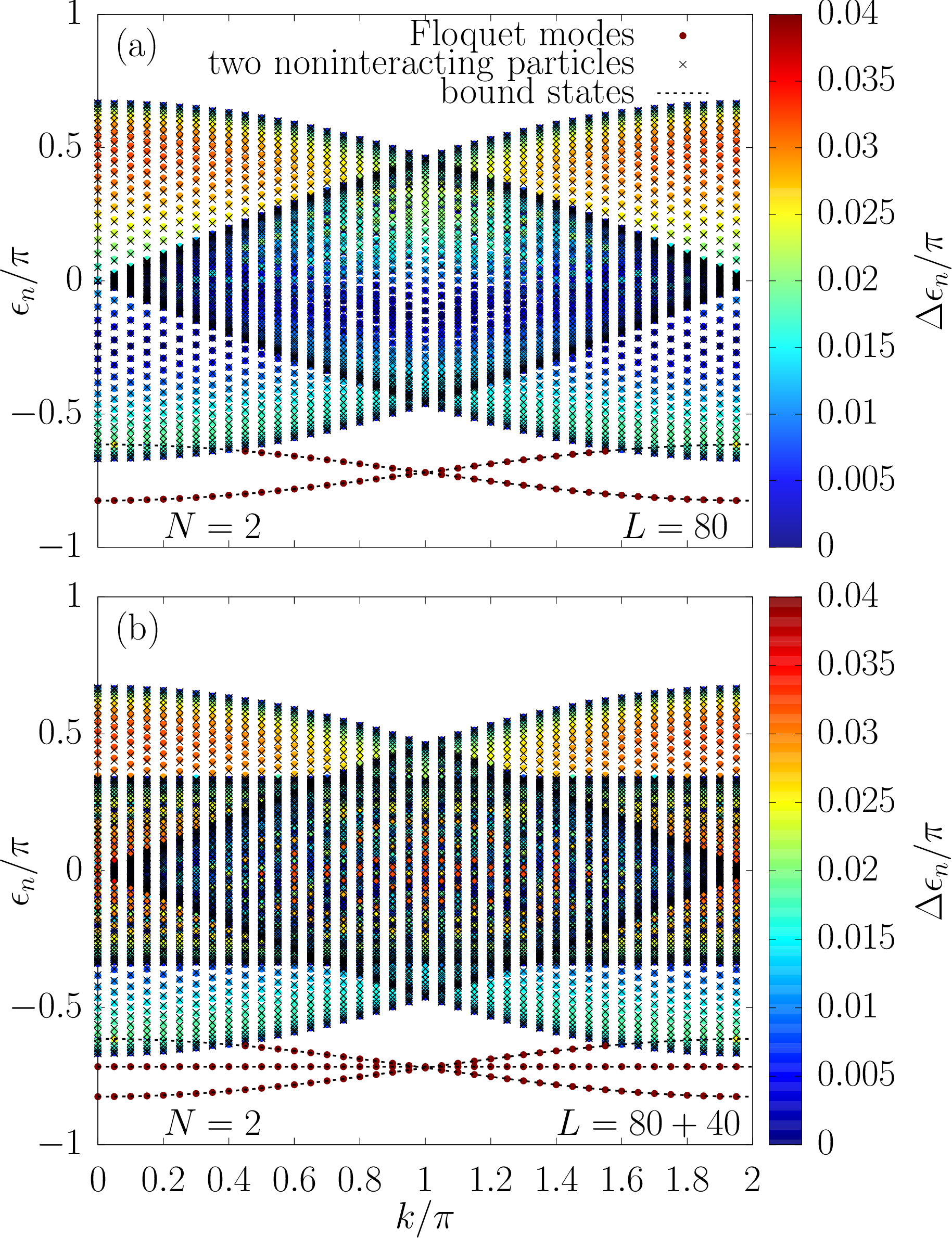}
 \caption{(a) Comparison of the actual dispersion of two-photon states for the integrable case $\theta'{=}0$ with no extra sites (dots) and the theoretical prediction for two separate noninteracting photons (crosses) and two bound photons (dashed lines). The color scale corresponds to the deviation between each dot and the closest cross.
 (b) Same as (a) but with added extra sites, while $\theta'{=}0$.  
 }\label{fig:special_states}
 \end{figure}

The sharp peaks in DOS can be attributed to the existence of special eigenstates with a relatively simple structure, which can be built by combining single-photon and two-photons states. Analytical expressions for the dispersions of a single photon or $N$-bound photons in the integrable (non-decorated) circuit are known~\cite{Aleiner2021}. Adding the decorations with $\theta^\prime{=}0$ results in an additional zero-energy band in the single-photon dispersion, since all the photons in the extra sites are frozen. This means that, for example, single-photon and two-photon eigenstates are still present in the three-photon spectrum at $\theta^\prime{=}0$, since we can just move the remaining photons to the decorations, where they will have zero energy.

In Fig.~\ref{fig:special_states}(a), we compare the actual two-photon Floquet spectrum (dots) with the states constructed from two single-photon states (crosses). The color scale represents the deviation from the nearest analytically constructed state. The agreement is remarkably good, which is not surprising given that the system is very dilute and only nearest-neighbour interactions are present. The two bands at the bottom of the plot are two-photon bound states, which are also in agreement with analytical expressions from Ref.~\cite{Aleiner2021}. Figure~\ref{fig:special_states}(a) is for the integrable case with no extra sites. Adding the decorations leads to the appearance of two additional bands, see Fig.~\ref{fig:special_states}(b) and compare with (a). The first one is a bound-state band, which corresponds to one photon in the main chain and another in adjacent decoration and is completely flat. The second one is a wider band of single-photon states corresponding to one photon inside and the other in a non-adjacent decoration. This wider band is centered around zero and has high DOS on its edges, which coincides with the peaks around $\pm\pi/3$ in three-photon DOS from Fig.~\ref{fig:dos}(a). Another smaller peak in DOS around $-0.75\pi$ comes from the flat band of two bound photons. 

We can conclude from the previous discussion that the three-photon DOS is strongly influenced by special single- and two-photon eigenstates. This effect is not so prominent in DOS for $4$ or more photons, likely because the number of special states is much smaller compared to the total Hilbert space size. In Appendix~\ref{appendix:excitation_density} we quantify this and show that the proportion of special states for a fixed photon number $N$ becomes asymptotically independent of the system size $L$. However, the saturation value still strongly depends on $N$, e.g., the special states comprise as many as 70\% of all states for $N{=}3$ but only 1\% for $N{=}8$. 

\begin{figure}[bth]
 \includegraphics[width=0.425\textwidth]{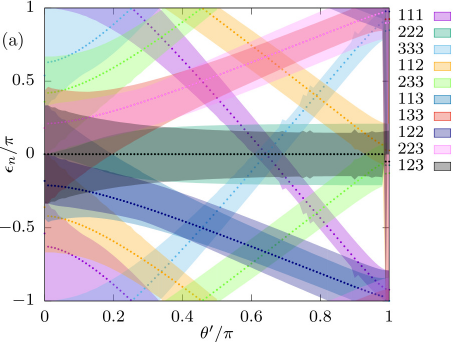}\\
\includegraphics[width=0.43\textwidth]{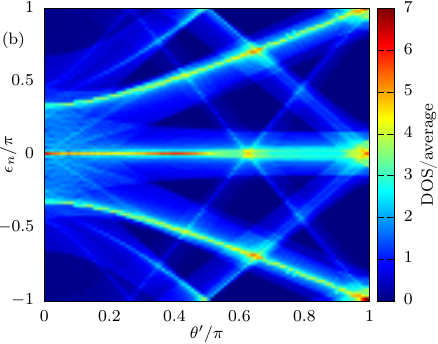}
 \caption{(a) Bands of special states constructed from numerically-obtained single-photon bands for $\theta'\in[0,\pi]$. Here we consider the case of $N=3$ separate photons. The bands are denoted by the number of photons in the first (1), second (2) and third (3) single-photon band. (b) Corresponding DOS. The bright regions can be related to the peaks in Fig.~\ref{fig:dos}.
 }\label{fig:special_states_theta}
 \end{figure}

 This analysis can now be extended to finite values of $\theta'$. The analytical expressions for the single-photon dispersion are not available in this case, but can be easily numerically computed for different coupling strengths $\theta'$. There are still three different bands, since each unit cell contains three sites. This numerical data can be used to construct three-photon bands, which correspond to three separate particles. This is a good approximation in a dilute system, even with non-zero interaction strength. In this way we obtain ten different bands, e.g., all three photons in the first band (denoted by 111), two photons in first and one in second (112), one photon per each band (123), etc. The dependence of these bands of special states on $\theta'$ is shown in Fig.~\ref{fig:special_states_theta}(a). As $\theta'$ is increased, the bands move and cross each other. 

\begin{figure*}
 \includegraphics[width=0.49\textwidth]{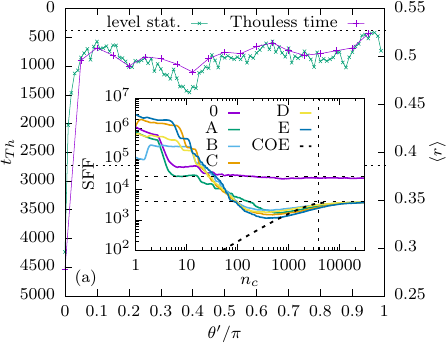}
 \includegraphics[width=0.49\textwidth]{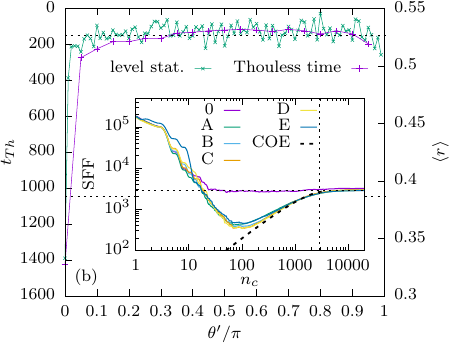}
 \caption{
 (a) Comparison of the level statistics ratio $\langle r\rangle$ and the Thouless time extracted from the spectral form factor (SFF) for $N=3$ photons, $L=60+30$.
 Inset: SFF time series corresponding to the main plot for several values of $\theta'$ [the labels A--E are defined in Fig.~\ref{fig:level_statistics}(a)]. The data was averaged over $\beta$ and smoothened by a moving average (see text).  Note the logarithmic scale on both the $x$ and $y$ axis.
 The horizontal dashed lines mark the saturation values, $\mathcal{H}$ and $\mathcal{H}+n_0^2$, while the vertical line at $n_c=\mathcal{H}$ is the Heisenberg time. The dashed black curve is the COE prediction for the linear ramp. 
 The agreement with COE becomes better as $N$ increases.
 (b) Same for $N=5$ and $L=14+7$, which shows a much clearer linear ramp in the SFF and better agreement with the random matrix theory at late times.
 }\label{fig:sff}
 \end{figure*}
  
 The DOS is typically higher near the edges of the bands, so we expect the DOS to be amplified when two bands cross. The edges of $222$ and $123$ bands overlap around $\theta'{=}0.35\pi$, which is where the level spacing ratio deviates the most from $\langle r\rangle_\text{COE}$. Several other bands also overlap around this point. The DOS plot for the special bands shown in Fig.~\ref{fig:special_states_theta}(b) roughly corresponds to the peaks in Fig.~\ref{fig:dos}(a). Therefore, as in the $\theta'{=}0$ case, the peaks in DOS at $\theta'{\neq}0$ are also explained by the special states which comprise a large proportion of the Hilbert space in systems with smaller numbers of photons, such as $N{=}3$. However, it is not obvious from Fig.~\ref{fig:special_states_theta}(b) in which $\theta'$ regions the level statistics deviates the most from values expected in chaotic systems. In particular, there are three very prominent peaks around $\theta'{=}0.65\pi$, where the level spacing distribution is actually very close to COE. We conjecture that these peaks are not narrow enough to lead to a sufficient number of degeneracies that could affect the level statistics. One might expect that the specially constructed states are a better approximation for a non-interacting system and that the $\langle r\rangle (\theta')$ dependence would look different at smaller values of the interaction strength $\phi$. This, however, is not the case, as shown in Appendix~\ref{appendix:parameters}, where it can be observed that the level statistics barely changes with $\phi$.

\subsection{Spectral form factor}\label{sec:sff}

The level statistics quantities considered above derive from the properties of eigenvalues of the Floquet unitary, hence they describe the behavior of the system at late times. In order to gain information about intermediate times, we study the spectral form factor (SFF)~\cite{haake2001}:
\begin{equation}
    K(t)=\sum_{m,n}e^{it(\epsilon_n-\epsilon_m)},
\end{equation}
which is defined in terms of two-point correlations between Floquet quasienergies, $\epsilon_n$. As we set the time period of one unitary cycle to $T{=}1$, the time in the above equation is equal to the number of cycles, $t{=}n_c$. 

The SFF is known to behave differently in integrable and chaotic systems, see Refs.~\cite{torres2017dynamical, Bertini2018, Chan2018, Suntajs2020} for some recent examples. In both cases, the SFF behavior at short times is governed by microscopic details of the system and therefore it is non-universal. 
After this initial transient, in integrable systems (Poisson ensemble) the SFF stays approximately constant around the value equal to the Hilbert space dimension $\mathcal{H}$, $K_\text{P}(t)\approx\mathcal{H}$.
In non-integrable systems, it first reaches a global minimum and, around the Thouless time $t_\text{Th}$, it starts to grow approximately linearly, according to the predictions of random matrix theory, until it saturates at $\mathcal{H}$ around the Heisenberg time $t_\text{H}\sim\mathcal{H}$. The level statistics and density of states studied previously naturally pertain to the times of order $t_\text{H}$, where the discreteness of the Floquet quasienergy spectrum is resolved. 

The SFF is typically noisy and suffers from a lack of self-averaging~\cite{Prange1997, Schiulaz2020}. In order to smoothen its time dependence, we choose to average it over the flux through the ring $\beta$. This parameter does not qualitatively affect the level statistics, as confirmed in  Appendix~\ref{appendix:parameters}. 
Additionally, after averaging over $100$ values of $\beta\in[0,\pi]$, we also compute the moving average at each time point by taking into account the nearest $60$ points, which finally results in relatively smooth curves. The averaged SFFs for $N=3$ photons, $L=60+30$ sites and different values of $\theta'$ are shown in the inset of Fig.~\ref{fig:sff}(a). After an initial period of non-universal behavior, the SFF for $\theta'=0$ assumes an approximately constant value, confirming that the system is integrable. In contrast, a clear linear ramp followed by saturation emerges for all studied values of $\theta'>0$, consistent with broken integrability.
We note that the SFF for $\theta'{=}0$ saturates at a higher value than $\theta'{>}0$, where the plateau is exactly as expected at $\mathcal{H}$. The reason for this is a large number $n_0$ of zero modes in the integrable case, which increases the late-time value of the SFF to $\mathcal{H}+n_0^2$ (at $\theta'{=}0$).

Furthermore, the Thouless time $t_\text{Th}$ can be extracted from the SFF data. This time gives us the onset of the universal behavior described by random matrix theory (i.e., the linear ramp). The COE prediction for SFF in the time window $0<t<\mathcal{H}$ is~\cite{Mehta2004}
\begin{equation}
    K_\text{COE}(t)= 2t-t \ln(1 + 2t/\mathcal{H}),
\end{equation}
as shown by the dashed black curve in Fig.~\ref{fig:sff}. 
In theory, the Thouless time could be defined as the smallest time for which $K(t)=K_\text{COE}(t)$. However, since $K(t)$ is typically not smooth enough even after averaging, in practice we use the following criterion to determine the Thouless time~\cite{Colmenarez2022}
\begin{equation}
    \ln(K(t_\text{Th})/K_\text{COE}(t_\text{Th}))=0.4.
\end{equation}
The precise value of the filtering parameter $0.4$ is unimportant, as long as it is finite but not too small.
In Fig.~\ref{fig:sff}, we plot the extracted Thouless time together with the average level spacing ratio $\langle r\rangle$ for $N{=}3$ and varying $\theta'$. Interestingly, the two curves exhibit very similar features, which means that the previously observed deviations in the level statistics for $N{=}3$ leave an imprint in the thermalization properties of the system, i.e., thermalization occurs later in systems which are farther away from the non-integrable case.

The agreement of the SFF with the random matrix theory prediction $K_\text{COE}$ is not particularly good for $N{=}3$ photons. This supports our previous observation that the energy spectra in small photon number sectors have special properties, e.g., as seen in the oscillations in level statistics and non-monotonic DOS. The agreement with COE becomes better as the number of photons $N$ increases, as can be seen for $N=5$ in Fig.~\ref{fig:sff}(b). The number of zero modes at $\theta'{=}0$ is now much smaller than the Hilbert space dimension, so the dashed horizontal lines at $\mathcal{H}$ and $\mathcal{H}+n_0^2$ are visually indistinguishable. There is also less variance in $K(t)$ curves for different values of $\theta'{>}0$, which is reflected in the almost constant value of the extracted Thouless time, as shown in the same figure. This is in line with the level spacing ratio $\langle r\rangle$, which shows no oscillation with $\theta'$ for this photon number but instead remains approximately constant around $\langle r\rangle_\mathrm{COE}$. 

 \begin{figure*}[bth]
  \includegraphics[width=0.99\textwidth]{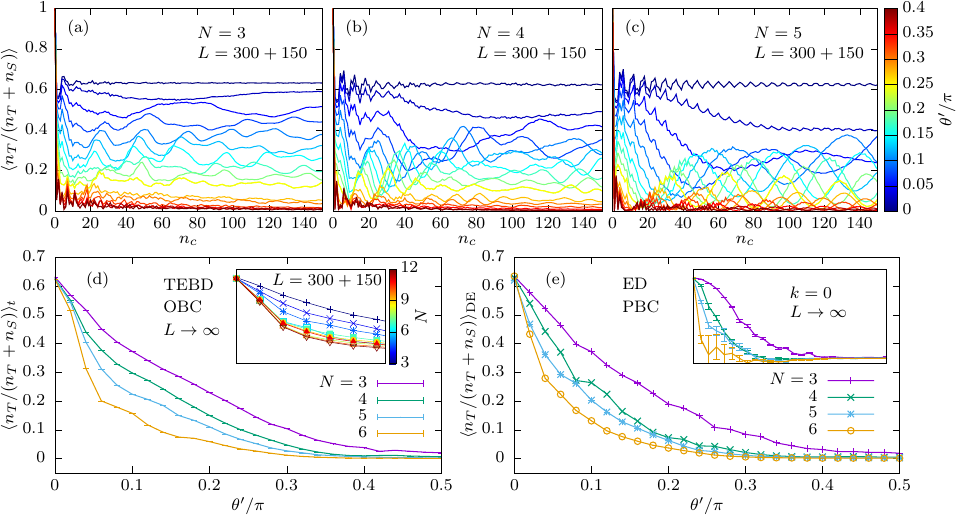}
 \caption{(a)-(c): Dynamics of the bound state probability (BSP) with $\theta{=}\pi/6$, $\phi{=}2\pi/3$, for $N{=}3$ photons (a), $N{=}4$ (b) and $N{=}5$ (c), at fixed system size $L{=}300{+}150$. Data is obtained using TEBD with bond dimension $\chi{=}256$ and open boundary conditions. 
(d) Time average of the BSP over 100 cycles using TEBD and extrapolated to  $L{\to}\infty$. We also averaged over two inequivalent initial states (\ref{eq:initial_state}). 
Inset: Average BSP for different photon numbers $N$ at fixed $L{=}300{+}150$.
Data was averaged over 150 cycles and obtained using TEBD with bond dimension $\chi=320$ for $3\leq N \leq 12$, $\theta=\pi/6$, $\phi=2\pi/3$
and varying $\theta^\prime$. 
 (e)~Diagonal ensemble prediction for the probability to remain in a bound state, averaged over two possible initial configurations. Data in this panel is obtained using ED with PBCs, on system sizes $N{=}3$, $L{=}30{+}15$ (with Hilbert space dimension $\mathrm{dim}{=}14189$); $N{=}4$, $L{=}20{+}10$ ($\mathrm{dim}{=}27404$); $N{=}5$, $L{=}14{+}7$ ($\mathrm{dim}{=}20348$); $N{=}6$, $L{=}12{+}6$ ($\mathrm{dim}{=}18563$).  
 Inset: Extrapolation to infinite system size for a translation invariant initial configuration with $k{=}0$ momentum.
 }\label{fig:BSP_dynamics}
 \end{figure*}

 \section{Bound states}\label{sec:bound_states}

Thus far, we have focused on generic aspects of thermalization at the level of the entire Floquet spectrum of the decorated XXZ circuit in Eq.~(\ref{eq:Uf}). However, one of the motivations behind the experiment~\cite{Google} to study this particular model is the fact that its integrable version hosts a special class of  ballistically propagating bound states. While such bound states are here protected by integrability, they represent only a fraction of all eigenstates and therefore it is not hard to imagine that they may persist, due to some other protective mechanism, after integrability is broken. We now examine in detail the stability of such states after decorating the circuit to break its integrability. 

In the Ising limit of the Hamiltonian version of the XXZ model, $J_z\gg J$, an $N$-particle bound state corresponds to $N$ adjacent spins being flipped~\cite{Ganahl2012}, 
\begin{equation}\label{eq:initial_state}
\lvert 000..\underbrace{1...1}_N..000\rangle.
\end{equation}
Even far from the Ising limit when $J_z\geq J$, the behavior of an $N$-particle bound state can be understood by starting from such an initial state, which is no longer an eigenstate. The same is true for the Floquet XXZ circuit, with the Hamiltonian Ising limit corresponding to $\phi\gg \theta$. 

Starting from the initial state (\ref{eq:initial_state}), the ``bound state probability'' (BSP) after $n_c$ cycles is given by
\begin{equation}\label{eq:BSP}
    \mathrm{BSP} = \frac{n_T}{n_T+n_S}, 
\end{equation}
where $n_T= \langle \psi(n_c) | \sum_j\prod^{j+(N-1)}_{i=j} \hat{n}_i |\psi(n_c) \rangle$ is the probability of finding photons in $N$ adjacent sites, where the indices $i$ and $j$ label the sites on the main chain. Conversely, the probability of any other $N$-photon configuration was denoted $n_S$. The BSP defined in this way was experimentally measured in Ref.~\cite{Google} and was found to gradually decay over time even at $\theta'{=}0$. This decay was due to experimental imperfections rather than an intrinsic property of the model. For an ideal implementation of the XXZ circuit, the BSP drops rapidly before fluctuating around a steady finite value, as will be shown below.  However, once integrability breaking terms are introduced into the Floquet circuit, there is no requirement for the $N$-photon bound states to continue to be stable at late times. Below, we will focus on understanding the effect of integrability breaking on BSP dynamics using TEBD simulations implemented in  iTensor~\cite{itensor}. Subsequently, we will show that other observables can reveal a signature of bound states by probing the memory of the initial configuration in Eq.~(\ref{eq:initial_state}) as the system evolves in time. Finally, we will interpret these results by examining the structure of the Floquet modes, in particular their overlap on the initial state in Eq.~(\ref{eq:initial_state}). 

\subsection{Dynamics of bound state probability}

The BSP dynamics for $N{=}3$, $4$ and $5$ photon bound states is presented in Figs.~\ref{fig:BSP_dynamics}(a)-(c) for various  strengths of the integrability-breaking $\theta^\prime$.  By increasing  $\theta^{\prime}$ from $0$ to $\pi/2$, the decorations become more strongly coupled to the main chain and the bound states are eventually destroyed. However, at intermediate $\theta^\prime$ the BSP does not appear to decay to zero, even after many cycles. This is true even when $\theta^\prime$ is comparable in size to the natural energy scale along the chain, $\theta^\prime\approx \theta$. For larger bound states, $\theta^\prime$ introduces large, slow oscillations into the BSP  that are independent of system size. The origin of these oscillations will be explained in Sec.~\ref{subsec:floquet}. 
 
Typically, an infinitesimally small perturbation is sufficient to destroy integrability in the thermodynamic limit $L{\rightarrow} \infty$ and infinite time limit $t{\rightarrow} \infty$. We access these limits by extrapolating the numerical data for the BSP via two methods: time-averaging the TEBD results and evaluating the diagonal ensemble predictions from ED data. The latter directly takes the $t{\to}\infty$ limit by assuming that the off-diagonal elements of the density matrix average out to zero~\cite{RigolNature, DAlessio2014}. These two methods have different advantages and limitations. While the TEBD method allows us to study  dynamics in very large systems, these simulations become computationally more expensive as the evolution time increases, which limits the total number of cycles.
On the other hand, the diagonal ensemble prediction provides information about the BSP at infinite time, however, it requires a computation of the complete eigenspectrum using ED, which limits the maximal system size. The total Hilbert space size is constrained by the amount of RAM available for diagonalization, while our implementation relies on 128-bit integers to represent basis configurations, which limits the maximal number of sites $L\leq 128$, irrespective of the photon number $N$. In principle, the latter restriction can be lifted using a more flexible encoding of the basis states, at the cost of sacrificing some of the computation efficiency.

For the TEBD time average of the BSP, we consider $100$ cycles between cycle $n_{c}=20$ and cycle $n_{c}=120$ for a variety of system sizes ranging from $L=20+10$ to $L=300+150$. In this way, we exclude the data at very short times which may be impacted by non-universal effects. By fitting the average BSP at each system size according to $\mathrm{BSP}(L,\theta^\prime)=\alpha(1/L)+\mathrm{BSP}_{\infty}$ we extrapolate to $L\rightarrow\infty$ and obtain the result plotted in Fig.~\ref{fig:BSP_dynamics}(d). This procedure was repeated for several photon numbers $N$. In each case, the initial state was chosen according to Eq.~\eqref{eq:initial_state}, which is not translation-invariant. As discussed in Sec.~\ref{sec:model}, there are two such inequivalent configurations and our results are averaged over both. We find that the bound states are robust for a finite range of $\theta^\prime$ which decreases as the size of the bound states increases.  
 
We also address how the robustness changes as the bound states increase in size, but continue to be dilute relative to the total system size, $N/L\ll 1$.  We calculate the BSP for bound states between sizes $N=3$ and $N=12$, averaged over $n_c=150$ cycles, to find $\textnormal{BSP}(N,\theta^\prime)$. 
The results for a fixed number of sites $L=300+150$ are plotted in the inset of Fig.~\ref{fig:BSP_dynamics}(d), where it can be seen that the $\textnormal{BSP}(N,\theta^\prime)$ curves are starting to converge for larger values of $N$.
These results suggest that large but dilute bound states continue to be robust. 

The diagonal ensemble results, which directly access the infinite time limit $t{\rightarrow} \infty$ of the BSP, can be seen in Fig.~\ref{fig:BSP_dynamics}(e). These results are consistent with the extrapolated TEBD results, suggesting small bound states are robust for a finite range of $\theta^\prime$. In particular, the $N{=}3$ bound states appear to be robust up to values of the integrability breaking that are comparable to the on-chain hopping terms. As $N$ becomes larger the bound states appear to become less robust but both Fig.~\ref{fig:BSP_dynamics}(e) and Fig.~\ref{fig:BSP_dynamics}(d) suggest the $N{=}4$, $N{=}5$ and $N{=}6$ bound states are robust for a finite range of $\theta^\prime$.  

For our diagonal ensemble calculations in Fig.~\ref{fig:BSP_dynamics}(e) we also averaged over two inequivalent initial configurations (\ref{eq:initial_state}). Since these states break translation invariance, we have to work in the full Hilbert space and therefore cannot obtain enough data points for reliable system-size scaling. However, if we form a translation-invariant initial state, we can restrict to the $k{=}0$ momentum sector and reach much larger system sizes. This allows us to extrapolate the diagonal ensemble value for BSP to $L{\rightarrow}\infty$, as shown in the inset of Fig.~\ref{fig:BSP_dynamics}(e). These results suggest that the bound states of $N{=}3$, $4$ and $5$ photons are robust at moderate values of  $\theta'$, which are clearly in the non-integrable regime according to the level statistics in Fig.~\ref{fig:level_statistics}, even in the infinite time and infinite size limit. 

Although the results for translation-invariant and non-invariant initial states in Fig.~\ref{fig:BSP_dynamics}(e) and inset are qualitatively similar, there are also some minor differences. Most notably,  the BSP decays quadratically with $\theta'$ for the symmetric state and linearly for the non-symmetric initial state, suggesting that symmetric states are more robust to integrability breaking. This is not surprising, given that Floquet modes have well defined momenta due to the overall symmetry of the system, and therefore a translation symmetric state can have a higher overlap with a single mode than a non-symmetric one. Nevertheless, these differences appear to rapidly diminish with an increase in $N$.

\subsection{Memory of the initial state}

The BSP is not the only local observable that reveals the unusual behavior of the bound initial states at finite $\theta^\prime$. Persistent nonthermalizing behavior can also be seen in the site occupation, $n_i=\langle \hat{n}_i\rangle$. Since the integrability breaking decorations make up one third of the total sites on the chain, we would expect a third of the photons to be located on them after a short time when the system has sufficiently thermalized. Instead, we find that this is only true at larger $\theta^\prime$ ($\gtrsim\pi/3$). The fraction of photons located on the decorations as $\theta^\prime$ is varied 
shows very similar behavior for bound states of different sizes. 

 \begin{figure}[bt]
 \includegraphics[width=0.45\textwidth]{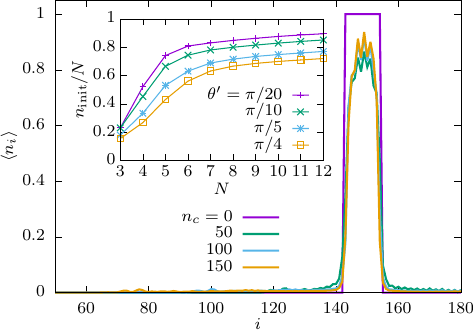}\\
 \caption{Site occupation of a $N{=}12$ photon  bound state after some number of cycles indicated on the legend. Data is obtained by TEBD for at system size $L=300+150$, bond dimension $\chi=320$ for $\theta=\pi/6$, $\phi=2\pi/3$ and  $\theta^\prime=\pi/4$.
 Inset: $n_{\mathrm{init}}/N$ for bound states of sizes $N=3$ to $N=12$ at different values of $\theta^\prime$. Data is obtained by TEBD for 150 cycles and the same parameters as in the main plot.
  }\label{fig:CentOcc}
 \end{figure}

In the integrable case, larger bound states propagate more slowly due to their small group velocity~\cite{Aleiner2021,Google}. This behavior appears to persist in the nonintegrable model. A large fraction of photons in the bound state remain in the vicinity of their initial sites even after many cycles. In Fig.~\ref{fig:CentOcc}, we show $n_i(n_c)$ for an $N{=}12$ bound state, demonstrating this robust nonthermalizing behavior. We can quantify this by calculating $n_{\mathrm{init}}=\sum_{i \in \mathrm{initial\, sites}}n_i$, the occupation of all the sites initially occupied by a photon. The average of $n_{\mathrm{init}}$ for different size bound states can be seen in the inset of Fig.~\ref{fig:CentOcc}. From this perspective, the bound states appear to grow more robust as they increase in size.
This at first seems to be in contradiction with the BSP results from Fig.~\ref{fig:BSP_dynamics}. However, the BSP measures the overlap with the $N$-photon bound state as a whole, while $n_{\mathrm{init}}$ also captures the case when the bound state loses photons from the edges while its core stays robust and does not move away significantly from its initial position. This is precisely what happens for large-$N$ bound states. Since we are considering hardcore bosons, the photons from the middle of the bound state can only hop to the decorations and back, while the photons at the edge can move further away along the chain and become detached from the rest.

 \begin{figure*}
 \includegraphics[width=0.98\textwidth]{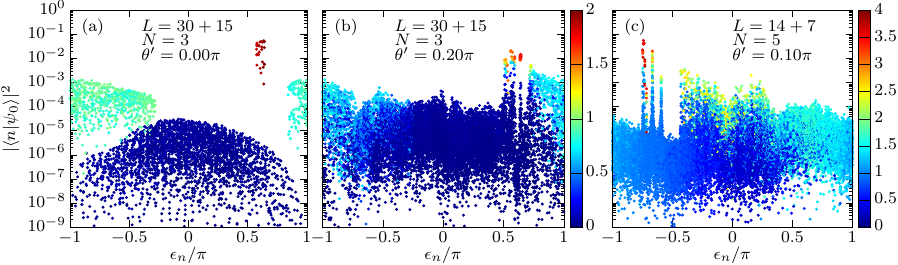}
 \caption{Overlap of the initial state with the eigenstates of the one-cycle evolution operator. (a)-(b): $N=3$ photons in system size $L=30+15$ for $\theta'=0$ and $\theta'=0.2\pi$, respectively. (c) $N=5$ photons in system size $L=14+7$ with $\theta'=0.1\pi$. The color scale is the number of pairs of adjacent occupied sites.  
 }\label{fig:overlap}
 \end{figure*}

\subsection{Floquet modes}\label{subsec:floquet}

 In order to understand the robustness of bound states to integrability breaking, we compare them with the Floquet modes of the system, i.e., the eigenstates  of $\hat{U}_F$. In Figs.~\ref{fig:overlap}(a)-(b) we plot the overlap of the all the eigenstates with the initial bound state for $N{=}3$ photons, contrasting the integrable case with that of $\theta^\prime=0.2\pi$. The color scale represents the number of pairs of neighboring occupied sites for each state. For example, the value of 2 corresponds to a three-photon bound state, 1 to two neighboring and one separate photon, while 0 implies three separate photons. The integrable case, $\theta'=0$, in Fig.~\ref{fig:overlap}(a) displays three separate sectors, one of which contains the bound states (red points). The energies of these sectors overlap, but there is no mixing between the states. The overlapping energies are a consequence of the periodic Floquet spectrum. In the Hamiltonian XXZ model these sectors are separated by energy gaps.  As $\theta'$ is increased, the system becomes non-integrable and the sectors start to mix. At $\theta'=0.2\pi$, Fig.~\ref{fig:overlap}(b), the bound states have almost merged with the bulk, but still remain visible. This is no longer the case after $\theta'=0.3\pi$, which is consistent with the results of Figs.~\ref{fig:BSP_dynamics}(d)-(e) where it was shown that the bound states are robust only up to this point. However, the bound states survive at values of $\theta'$ which are clearly in the non-integrable regime from the point of view of the level statistics, recall Fig.~\ref{fig:level_statistics}(a).
 
 We have also repeated these calculations for a translation-invariant initial state in the $k{=}0$ momentum sector. The main difference is the reduction in number of bound Floquet modes. There are two such states in the $k{=}0$ sector and their energies and overlaps with the initial state show almost no dependence on the system size $L$. This is related to the fact that the BSP does not significantly decay with the system size.

The overlap plots for larger numbers of photons display similar features to $N{=}3$. At $\theta'{=}0$, there are $N$ separate sectors which are defined by the number of pairs of adjacent photons. The bound Floquet modes slowly mix with the other sectors as the coupling to the decorations in increased. The case of $N{=}5$ and $\theta'{=}0.1\pi$ is shown in Fig.~\ref{fig:overlap}(c). Here we see three prominent towers of states, which are related to the oscillations in the BSP in Fig.~\ref{fig:BSP_dynamics}(c). These oscillations can be attributed to photons from the bound state hopping onto the extra sites and back.
The towers appear as soon as $\theta'{\neq}0$ and persist until approximately $\theta'{\approx}0.3\pi$. The distance between the towers and therefore the oscillation frequency depends approximately linearly on $\theta'$.
Moreover, the shape and height of the towers depend on the number of decorations attached to the initially occupied sites. 

Intriguingly, we find that the towers are more prominent for the five-photons initial bound state with two decorations on the second and fourth site [Fig.~\ref{fig:overlap}(c)] than for the state with three decorations on the first, third and fifth site. Upon closer inspection, some towers of high-overlap states can also be discerned for $N{=}3$ in Fig.~\ref{fig:overlap}(b). However, they are not as well differentiated as for $N{=}5$, and do not appear to be equally spaced in energy, which is the reason why the oscillations in BSP at $\theta'{=}0.2\pi$ are irregular, see Fig.~\ref{fig:BSP_dynamics}(a). Interestingly, the towers do not become better resolved with increasing the photon number and the case of $N{=}5$ actually features the sharpest towers and corresponding oscillations in BSP and various local observables, such as the number of photons in the extra sites. As a side note, similar looking towers of states are often found in systems that host ``quantum many-body scars''~\cite{Serbyn2021, MoudgalyaReview, ChandranReview}, however it is not clear whether similar physics occurs in the present case.

\section{Finite density of excitations and other decoration patterns}\label{sec:other}

Up to this point, we have mostly focused on the experimental setup of Ref.~\cite{Google}, restricting to the case with integrability-breaking decorations on every other site and small numbers of photons $N{\leq}6$. In this Section, we consider other cases that were not studied previously. In particular, we will now fix the filling factor instead of fixing the total photon number, which will allow us to investigate convergence to the thermodynamic limit by growing both the system size and the number of excitations, as conventionally done in the literature. Moreover, we will explore other decoration patterns, including those with decorations on every $n$-th site where the system still retains translational invariance, as well as completely random patterns which break all the symmetries.

\subsection{Fixed filling factor}

When we extrapolate systems with small but fixed photon numbers to the infinite number of sites, they become infinitely dilute. By contrast, the thermodynamic limit is conventionally taken by keeping the \emph{density} constant. Thus, we introduce the filling factor  $\nu=N/L_\mathrm{sites}$ and study properties of our circuit as both $N$ and $L_\mathrm{sites}$ are simultaneously increased such that $\nu$ remains constant. 

\begin{figure}[bth]
 \includegraphics[width=0.45\textwidth]{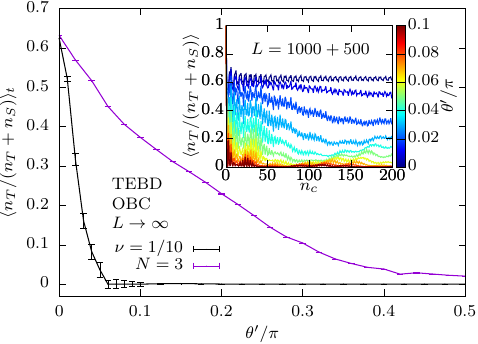}\\
 \caption{
 Time-averaged BSP extrapolated to infinite system size. Filling factor is fixed to $\nu{=}N/L_\mathrm{sites}{=}1/10$. For comparison, we also replotted the case of fixed $N=3$ from Fig.~\ref{fig:BSP_dynamics}.  Inset: BSP dynamics for $L{=}1000+500$ and $N{=}100$.
 All data is obtained by TEBD with bond dimension $\chi{=}256$.
  }\label{fig:fixed_nu}
 \end{figure}

As done previously for fixed $N$, we average the BSP (\ref{eq:BSP}) over a certain number of cycles and investigate its dependence on $\theta'$. In particular, the averaging was done between cycles $n_c{=}100$ to $n_c{=}200$, after the initial drop in bound state survival. The time-averaged value is then extrapolated to the thermodynamic limit, while keeping $\nu$ fixed, using a quadratic fit in $1/L$, 
see Fig.~\ref{fig:fixed_nu}. Here we set the filling factor to $\nu=1/10$, but the results for other $\nu$ values are similar. In contrast to dilute systems with fixed photon numbers, the average BSP now drops to very small values already at $\theta'\approx 0.05\pi$. 
Due to a very slow decay rate of the BSP for very weak integrability breaking $\theta'\lesssim 0.05\pi$, we expect Fig.~\ref{fig:fixed_nu} to provide only an upper bound, as the extrapolated value of the BSP would likely be smaller with an access to a longer time window. However, the TEBD calculations become significantly more time consuming with increasing number of cycles, which is a limiting factor in very large systems.

In the inset of Fig.~\ref{fig:fixed_nu} we show the evolution of BSP for $N{=}100$ photons in system size $L{=}1000{+}500$,  and several values of $\theta'$ obtained using TEBD. Unlike the case of fixed but small photon numbers, the large-$N$ bound states are less resilient to integrability breaking by coupling to the extra sites. The BSP quickly decreases with the number of cycles, as can be seen in the inset, where we only show the relatively small coupling strengths $\theta'\in[0,0.1\pi]$. The extrapolated values in Fig.~\ref{fig:fixed_nu} point to the conclusion that very small photon number sectors we studied up to this point have unconventional properties, which are not shared by thermodynamically large sectors with non-zero filling factor $\nu$.

\subsection{Other decoration patterns}\label{subsec:patterns}

\begin{figure}[bth]
\includegraphics[width=0.49\textwidth]{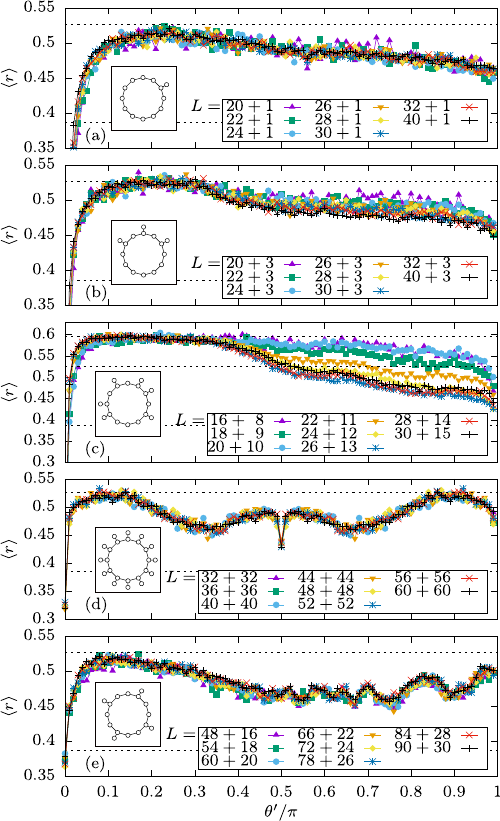}
 \caption{Level statistics for different decoration patterns (both translation-symmetric and non-symmetric). Data is for $N{=}3$ photons, $\theta{=}\pi/6$, $\phi{=}2\pi/3$, and various system sizes specified in legends. The horizontal dashed lines are expectations for the relevant ensembles, $\langle r\rangle_\mathrm{P}\approx0.386$, $\langle r\rangle_\mathrm{COE}\approx0.527$ and $\langle r\rangle_\mathrm{CUE}\approx0.597$. (a) One decoration on site 2. (b) Three decorations on sites 2, 4 and 6. (c) Random non-symmetric patterns. (d) Decorations on every site, in the momentum $k{=}0$ and inversion-symmetric $I{=}{+}1$ sector. (e) Decorations on every third site, $k{=}0$, $I{=}{+}1$ sector.
 }\label{fig:level_statistics_pattern}
 \end{figure}

Finally, we consider the level statistics and BSP dynamics for different patterns of decorations. In Fig.~\ref{fig:level_statistics_pattern} we show the average level spacing ratio $\langle r\rangle(\theta')$ for $N{=}3$ photons and five different types of decoration arrangements. The first one [Fig.~\ref{fig:level_statistics_pattern}(a)] is only a single decoration, while the second one [Fig.~\ref{fig:level_statistics_pattern}(b)] consists of three decorations attached to sites 2, 4 and 6. Both of these patterns break translation symmetry, which limits the system sizes we can reach. Unlike the level statistics in Fig.~\ref{fig:level_statistics}(a), where $\langle r\rangle$ was oscillatory for $N{=}3$, here we observe no such oscillations. However, $\langle r\rangle$ first increases to the COE value and then starts to decrease around $\theta'\approx0.3\pi$. 
The decrease does not happen for larger photon numbers, with the $\langle r\rangle(\theta')$ curve becoming flat already at $N{=}4$. Thus, we conclude that $N{=}3$ displays anomalous level statistics properties, irrespective of the decoration pattern.  
We have also examined the DOS for these other patterns of extra sites (not shown). The DOS distribution for the cases of one and three decorations differs from  Fig.~\ref{fig:dos}(a) in that it loses the sharp peaks, but the overall shape stays approximately the same and becomes flatter with increasing number of photons. The peaks likely disappear because the new decoration pattern is no longer translation-invariant, and the eigenstates which were previously degenerate are no longer related by symmetry, hence they generally have different energies.

Note that the $\langle r\rangle$ plateau is more pronounced and closer to the COE value for three decorations compared to a single decoration. This trend continues as we add more decorations. In Fig.~\ref{fig:level_statistics_pattern}(c) we show several random patterns where the number of extra sites is kept equal to half the number of sites inside the ring. We again observe similar behavior, with an initial plateau followed by a decrease in $\langle r\rangle$. However, the plateau is now at the CUE value $\langle r\rangle_\mathrm{CUE}\approx0.597$ instead of $\langle r\rangle_\mathrm{COE}\approx0.527$. As explained in Sec.~\ref{subsec:symmetry}, this is due to all of the studied random patterns breaking inversion symmetry, unlike the previous cases of one and three decorations. 
It might seem surprising that $\langle r\rangle$ is not monotonic with system size in Fig.~\ref{fig:level_statistics_pattern}(c), but this is simply due to choosing completely different patterns for each system size and could be avoided by averaging over several random patterns for each $L$. The DOS distribution for the cases in Fig.~\ref{fig:level_statistics_pattern}(c) again has no peaks for $\theta'{\neq}0$ and is noticeably flatter than the previously considered patterns (data not shown).

We have also considered two examples of periodic patterns, one with decorations attached to every site of the main ring [Fig.~\ref{fig:level_statistics_pattern}(d)] and the other with decorations on every third site [Fig.~\ref{fig:level_statistics_pattern}(e)]. The first case is invariant to translations by two sites and inversion which swaps the $i$-th site (decoration) with $(L_\mathrm{sites}-i)$-th [$(L_\mathrm{decor}-i)$-th], so these symmetries must be resolved in order to obtain the correct level statistics. We note that the full system is not inversion symmetric for the usual periodic pattern with decorations on every other site, even though the arrangement of decorations itself is. This is due to first applying the fSim gates on odd bonds and then on even bonds, Eq.~\eqref{eq:Uf}. There is no reflection axis which simultaneously preserves both the decoration pattern and the order of even and odd fSim gate layers.
Similar to previous results in Fig.~\ref{fig:level_statistics}(a), for $N{=}3$ we again observe deviations from $\langle r\rangle_\mathrm{COE}$ at certain values of $\theta'$. However, the $\langle r\rangle(\theta')$ curve is now symmetric around $\theta'=\pi/2$ with an integrable point in the middle. This is similar to the case in Fig.~\ref{fig:level_statistics_param}(a) in Appendix~\ref{appendix:parameters}, where $\langle r\rangle(\theta')$ is symmetric around $\theta{=}\pi/2$. As can be seen from Eq.~\eqref{eq:fSim}, this value of the hopping amplitude corresponds to a photon moving to the neighboring site with probability $1$, so it is not surprising that this is a special case.

For the second periodic pattern [Fig.~\ref{fig:level_statistics_pattern}(e)],  the symmetries of the full system are translation by six sites and inversion which preserves this arrangement of decorations. In this case, we also observe oscillations in $\langle r\rangle(\theta')$, but the local minima and maxima are at different values of $\theta'$ compared to the other patterns.
As before, all oscillations disappear for $N{=}4$ or more photons. 
For all studied periodic patterns with decorations on every $n$-th site, the DOS still exhibits pronounced peaks for $N{=}3$ and to some extent for $N{=}4$.

 \begin{figure}[tb]
 \includegraphics[width=0.45\textwidth]{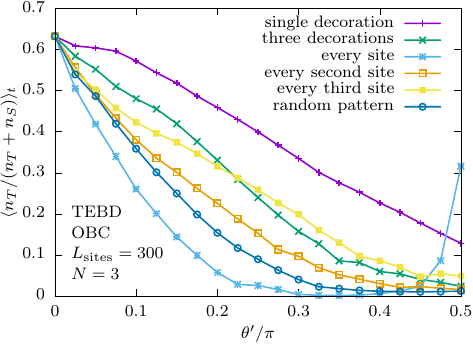}\\
 \caption{Time averaged BSP over 100 cycles for different decoration patterns in the $N=3$ excitation sector for $L_\mathrm{sites}=300$. The parameters given are $\theta=\pi/6$, $\phi=2\pi/3$ and $\chi=256$. 
 In the case of decorations on every second (third) site the results were averaged over two (three) non-equivalent initial bound-state configurations.
 The random pattern results were averaged over five different patterns with decorations on half of the sites, $L_\mathrm{decor}/L_\mathrm{sites}=1/2$.
  }\label{fig:dynamics_different_decs}
 \end{figure}

We have also investigated the robustness of bound states for various decoration patterns. In Fig.~\ref{fig:dynamics_different_decs} we plot the averaged BSP for $N{=}3$ photons, $L_\mathrm{sites}{=}300$ sites on the main chain averaged over $n_{c}{=}100$ cycles between cycles $n_{c}{=}50$ and $n_{c}{=}150$. For a decoration on every second site we average over the two possible $N{=}3$ photon initial states around the center of the chain. We perform a similar averaging over the three possible initial states for the case of a decoration on every third site. For randomly allocated decorations we instead average over 5 different random patterns with the initial bound state at the center of the chain. These results suggest the robustness of the bound states is more dependent on the density of decorations $L_\mathrm{decor}/L_\mathrm{sites}$ than on the actual pattern. The bound states survive for larger values of $\theta'$ when the number of decorations is smaller. The average BSP decays over a similar range of $\theta'$ for random patterns with $L_\mathrm{decor}/L_\mathrm{sites}{=}1/2$ and for the periodic case of the same site density. For the case of decorations on every site there is a peak at $\theta'{=}\pi/2$ which corresponds to the near integrable point as seen in Fig.~\ref{fig:level_statistics_pattern}(d). As before, the bound states become increasingly less robust as the number of photons grows.

 \begin{table*}
\begin{tabular}{ |c|c|c|c|c|c|c| }
\hline
$N$ & $L_\mathrm{decor}/L_\mathrm{sites}$ & pattern & robust bound states & level statistics & DOS & Figs. \\
\hline
\hline
 & 1 & & & & & \ref{fig:level_statistics_pattern}(d)\\  
 & 1/2 & & & & & \ref{fig:level_statistics}(a), \ref{fig:dos}\\ 
3 & 1/3 & periodic & \checkmark & oscillating $\langle r\rangle(\theta)$ &  
sharp peaks & \ref{fig:level_statistics_pattern}(e)\\
 & 1/4 & & & & &\\ 
 & 1/5 & & & & &\\
\hline
3 & single decoration & non-periodic & \checkmark & $\langle r\rangle_\mathrm{COE}$ plateau then decrease & no peaks & \ref{fig:level_statistics_pattern}(a)\\ 
 & three decorations & & & & & \ref{fig:level_statistics_pattern}(b)\\
\hline
3 & 1/2 & random & \checkmark & $\langle r\rangle_\mathrm{CUE}$ plateau then decrease & no peaks & \ref{fig:level_statistics_pattern}(c)\\
\hline
\hline
 & 1 & & & & & \\
4 and 5 & 1/2 & periodic & \checkmark & $\langle r\rangle_\mathrm{COE}$ for $\theta'\gtrsim0.05\pi$ & relatively flat & \ref{fig:level_statistics}(b), \ref{fig:level_statistics}(c), \ref{fig:dos}\\
 & 1/3 & & & & &\\
\hline
4 and 5 & single decoration & non-periodic & \checkmark & $\langle r\rangle_\mathrm{COE}$ for $\theta'\gtrsim0.05\pi$ & relatively flat & not shown\\ 
 & three decorations & & & & & \\
 \hline
 4 and 5 & 1/2 & random & \checkmark & $\langle r\rangle_\mathrm{CUE}$ for $\theta'\gtrsim0.05\pi$ & relatively flat & not shown \\
\hline
\hline
$\nu=1/10$ & 1/2 & periodic & $\cross$ & not computed & not computed & \ref{fig:fixed_nu}\\
\hline
\end{tabular}
\caption{Brief summary of different systems studied in this work. $N$ is the number of photons, 
$L_\mathrm{sites}$ is the number of sites on the main chain and $L_\mathrm{decor}$ the number of decorations, while $\nu=N/L_\mathrm{sites}$ is the filling factor. 
Periodic patterns are those with decorations on every single, second, third, fourth or fifth site.
}
\label{tab:summary}
\end{table*}

\section{Conclusions and discussion}\label{sec:discussion}

We have performed systematic classical simulations of the Floquet XXZ circuit on a 1D chain with integrability breaking decorations. This study was motivated by the recent Google experiment~\cite{Google} which realized the same model on a ring of superconducting qubits and investigated the dynamics of its bound states. Surprisingly, the bound states were observed to be resilient to integrability breaking perturbations in the form of extra qubits attached to the ring.  
We have analyzed the level statistics of this model and simulated the dynamics of bound states, confirming that some of these states indeed survive in certain parts of the non-integrable regime, even for an infinite number of qubits and at infinite time. In contrast to much previous work, the focus of Ref.~\cite{Google} and our own was on dilute systems containing \emph{few} excitations, which have rarely been discussed in the literature (see, however, Ref.~\cite{Zisling2021}). As we have demonstrated, such systems are amenable to classical simulations in  large numbers of qubits, providing useful benchmarks for future studies on improved quantum hardware. 

One of our most significant findings is that small but fixed photon number sectors show unusual properties in several respects. In particular, the robustness of bound states depends on the photon number, with larger states decaying more rapidly as the coupling to the integrability-breaking extra sites is increased. Moreover, the energy spectrum for $N{=}3$ photons has unusual level statistics, which deviates from the expectation for a chaotic system even for strong couplings to the integrability-breaking decorations. This was attributed to the presence of special eigenstates in the energy spectrum. These eigenstates have a relatively simple structure, which is related to one-photon and two-photon states, while the rest of the photons are located in the decorations. The proportion of such states is large enough only in sufficiently dilute systems. When the decoration pattern is periodic, some of these eigenstates are related by translation and are therefore degenerate in energy, which results in prominent peaks in the DOS and affects the level statistics. The deviations in level statistics were shown to leave an imprint in the dynamics of bound states by slowing down the thermalization. 

Additionally, we have investigated systems with constant filling factors and their extrapolation to the thermodynamic limit. Such systems are no longer dilute and our findings indicate that they do not support stable bound states when integrability is broken. This is in stark contrast with the bound states in very dilute systems with small photon numbers, such as the one studied in experiment~\cite{Google}. Moreover, we have explored other decoration patterns, including both periodic and non-periodic ones. A brief summary of all considered systems is given in Table~\ref{tab:summary}. Our calculations suggest that the peaks in the density of states disappear when the pattern is not periodic, which destroys the translation symmetry of the full system. This is likely a consequence of certain eigenstates no longer being degenerate. We also find that the inversion symmetry of the decoration patterns (or lack of it) influences the level statistics. In particular, inversion-symmetric patterns are consistent with COE and non-symmetric with CUE statistics in the chaotic regime. Deviations from these values were observed only for $N{=}3$ photons and were found to diminish as the number of photons is increased. However, our results do not indicate a link between the irregularities in level statistics and the robustness of bound states, although both properties are most prominent in dilute systems. For example, non-periodic decoration patterns result in level spacing ratios consistent with random matrix theory, implying that the integrability is indeed fully broken, while the few-photon bound states remain robust in that regime.

One advantage of classical simulations performed in this work is direct access to the system's properties at finite energy densities. Thus, the model considered in this work would be a useful platform for benchmarking quantum algorithms that target states at finite energy density~\cite{Banuls2021}. Moreover, it would be interesting to explore other models which host bound states, e.g., the chiral Hubbard model~\cite{Aleiner2021}, and investigate if such models exhibit similar behavior in relation to the density of excitations and integrability breaking by changing the geometry of the system, as described in this work. 

{\sl Note added:} After the completion of this work, we became aware of Ref.~\cite{Surace2023} which also studied the stability of bound eigenstates in the special case of $N{=}3$ photons and decorations on every second qubit. Based on perturbative arguments and the scaling of inverse participation ratio, Ref.~\cite{Surace2023} concluded that $N{=}3$ eigenstates slowly lose their bound state character in the $L{\to}\infty$ limit. Our finite-size scaling analysis above suggests that the bound state probability remains finite in this limit for $N{=}3$, however this cannot rule out the possibility of a much larger length scale, at which all dynamical signatures of bound states would ultimately disappear at infinite time.

\begin{acknowledgments}
We would like to thank Alexios Michailidis and Jean-Yves Desaules for useful discussions. We acknowledge support by the Leverhulme Trust Research Leadership Award RL-2019-015 and EPSRC Grant No. EP/R020612/1. Statement of compliance with EPSRC policy framework on research data: This publication is theoretical work that does not require supporting research data. This research was supported in part by the National Science Foundation under Grant No. NSF PHY-1748958.
A.H.~acknowledges funding provided by the Institute of Physics Belgrade, through the grant by the Ministry of Science, Technological Development, and Innovations of the Republic of Serbia. Part of the numerical simulations were performed at the Scientific Computing Laboratory, National Center of Excellence for the Study of Complex Systems, Institute of Physics Belgrade. 
\end{acknowledgments}

\appendix

\section{Continuous model} \label{appendix:XXZ}

For simplicity, here we assume a decoration pattern where one extra site is attached to every even site of the main ring with PBCs.
The continuous XXZ Hamiltonian which corresponds to the unitary circuit from Eqs.~\eqref{eq:fSim}-\eqref{eq:Uf} in the $dt\rightarrow 0$ limit is

\begin{equation}
\begin{aligned}\label{eq:xxz2}	
H_\mathrm{XXZ} &= \sum_{i=1}^{L_\mathrm{sites}} J(e^{i\beta}\sigma_+^i \sigma_-^{i+1}+e^{-i\beta}\sigma_-^i \sigma_+^{i+1})+J_z(\sigma_z^i \sigma_z^{i+1})\\
&+ \sum_{i=1}^{L_\mathrm{decor}}
J'(e^{i\beta}\sigma_+^{2i} \sigma_-^{e_i}+e^{-i\beta}\sigma_-^{2i} \sigma_+^{e_i})+J'_{z}(\sigma_z^{2i} \sigma_z^{e_i})\\
&+ \sum_{i=1}^{L_\mathrm{sites}} h_z(i)\sigma_z^{i}
+ \sum_{i=1}^{L_\mathrm{decor}} h_z^e\sigma_z^{e_i},
\end{aligned}
\end{equation}
where $e_i$ are the integrability-breaking extra sites attached to even sites $2i$.
The local field is $h_z(2i+1)=-2J_z$ on odd sites, $h_z(2i)=-3J_z$ on even sites and $h_z^e=-J_z$ on extra sites. Additionally, if we impose OBCs and even number of sites, the local field is $h_z(1)=-J_z$ on the first site and $h_z(L)=-2J_z$ on the last site. The corresponding unitary circuit parameters are  
$\theta=2Jdt$, $\phi=2J_zdt$, $\theta'=2J'dt$, and $\phi'=2J'_zdt$.
This continuous model can be easily generalized to an arbitrary decoration pattern by changing the local fields.

\begin{figure}[bth]
\includegraphics[width=0.45\textwidth]{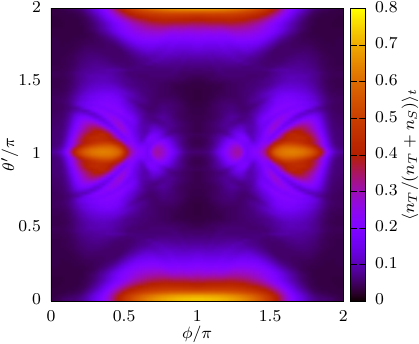}
 \caption{Bound state probability averaged over $n_c{=}50$ cycles and two different initial configurations. $L{=}40+20$, $N{=}3$, $\theta{=}\pi/6$, $\beta{=}0$, $\theta'\in[0,2\pi]$, and $\phi\in[0,2\pi]$. }\label{fig:scan}
 \end{figure}

 \section{Effect of parameters $\theta$, $\phi$ and $\beta$}\label{appendix:parameters}

Throughout this paper, we have mostly considered the parameters $\theta=\pi/6$, $\phi=2\pi/3$, and $\beta=0$, which were used in the experiment from Ref.~\cite{Google}. In this Appendix, we explore the effects of changing these parameters.
First we perform a scan of the parameter space over a range of interaction strengths $\phi$ and couplings to the extra sites $\theta'$ while keeping the hopping amplitude inside the ring fixed to $\theta/6$. The BSP for an initial state of $N{=}3$ bound photons, Eq.~(\ref{eq:initial_state}), averaged over the first $n_c{=}50$ unitary cycles, is shown in Fig.~\ref{fig:scan}. Here we also average over two non-equivalent initial configurations, although the plot looks very similar without averaging.  The results are symmetric to reflection around the $\phi=\pi$ and $\theta'=\pi$ axes, which is unsurprising as these reflections result only in minus signs in certain matrix elements from Eq.~\eqref{eq:fSim} [$\sin(2\pi-\theta')=-\sin(\theta')$ and $e^{i(2\pi-\phi)}=e^{-i\phi}$].

The region of interest is the bottom part of this figure, with intermediate $\phi$ and small $\theta'$. There is a minimal value of interaction strength required for the survival of bound states, $\phi_\mathrm{min}\approx\pi/3\approx2\theta$. This is in line with previous analytical results which state that the bound states exist for any momentum in the gapped regime $\phi{>}2\theta$~\cite{Aleiner2021}.
The maximal decoration coupling which supports the bound states is around $\theta'{=}\pi/3$ and does not significantly depend on $\phi$. There are also additional regions where the bound states remain stable, such as around $\phi{=}\pi/3$ and $\theta'{=}\pi$. However, the $\theta'{=}\pi$ line is a special case since there is no hopping from the main chain to the extra sites, see Eq.~\eqref{eq:fSim}. We therefore did not focus on these regions of the phase diagram.
We have also investigated the cases of $N{=}4$ or more photons in the initial bound state and these plots show similar features to $N{=}3$. The main difference is that the region with robust bound states shrinks in the $\theta'$ direction with increasing $N$, which is consistent with our results from Figs.~\ref{fig:BSP_dynamics}(d)-(e).

\begin{figure}[t]
\includegraphics[width=0.49\textwidth]{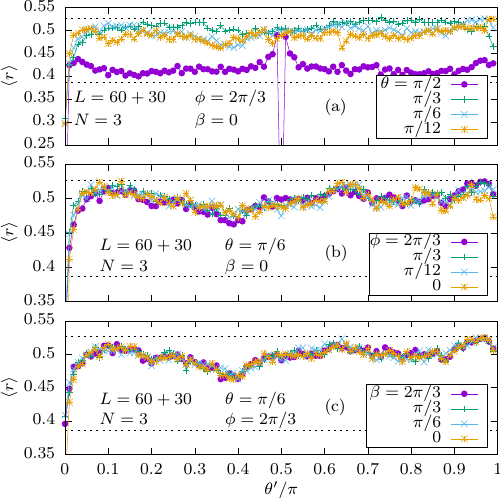}
 \caption{Level statistics for different values of parameter (a) $\theta$. (b) $\phi$. (c) $\beta$. The only parameter which significantly affects the results is the hopping amplitude $\theta$. We note that $\theta=\pi/2$ is a special case where the probability of hopping to the neighboring site is $1$ in each cycle, see Eq.~\eqref{eq:fSim}. It is therefore not surprising that this case is close to being integrable.
 }\label{fig:level_statistics_param}
 \end{figure}
 
In Fig.~\ref{fig:level_statistics_param} we plot the average level spacing ratio $\langle r\rangle(\theta')$ for several values of the parameters $\theta$, $\phi$ and $\beta$.  
As can be observed in Fig.~\ref{fig:level_statistics_param}(a), the oscillations in $\langle r\rangle(\theta')$ are visible for all values of the hopping amplitude $\theta$, but the exact positions of the local minima and maxima depend on $\theta$. The most distinctive case is $\theta{=}\pi/2$, where the level statistics is close to Poisson for most values of $\theta'$. According to  Eq.~\eqref{eq:fSim}, this is a special case where the hopping probability on the main chain is $1$ in each cycle.
Additionally, we note that the $\langle r\rangle(\theta')$ curve is the same for $\theta$ and 
$\pi-\theta$. 

In contrast to $\theta$, the other two parameters $\phi$ and $\beta$ have almost no effect on the level statistics.
The dependence on the nearest-neighbor interaction strength $\phi$ is shown in Fig.~\ref{fig:level_statistics_param}(b). This further supports the conclusion that the deviations from $\langle r\rangle_\mathrm{COE}$ are mainly a consequence of  
three separate photon
states whose energies do not depend on $\phi$. 
The magnetic flux through the ring $\beta$ has even less influence on the results, see Fig.~\ref{fig:level_statistics_param}(c). However, it does change the actual energy levels.
For this reason, the parameter $\beta$ was very useful for averaging the spectral form factor in Sec.~\ref{sec:bound_states}.

 \begin{figure}[b]
 \includegraphics[width=0.49\textwidth]{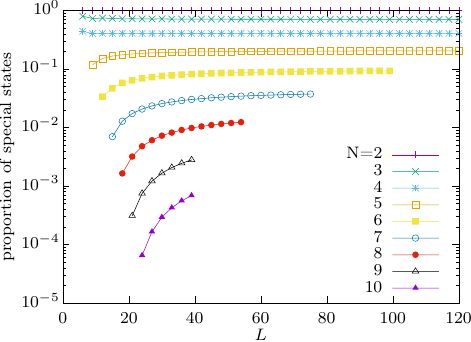}
 \caption{Proportion of special configurations for different photon numbers $N$ and system sizes $L=L_\mathrm{sites}+L_\mathrm{decor}$. The decorations are on every other site. These simple states can be constructed by combining single-photon and two-photon configurations with photons in the extra sites.
 }\label{fig:special_states_proportion}
 \end{figure}

\section{Density of excitations}\label{appendix:excitation_density}

 We have previously attributed the deviations from the chaotic level statistics and sharp peaks in DOS to the existence of special eigenstates in  Sec.~\ref{subsec:dos}. These features were more pronounced for small photon numbers such as $N{=}3$.
 In Fig.~\ref{fig:special_states_proportion} we show the proportion of some simple basis configurations in total Hilbert spaces of systems with different fixed numbers of photons $N$ and increasing number of sites $L$. In particular, these are the configurations with all photons outside the main chain, just one photon inside, two separate photons inside, and two bound photons inside. The special eigenstates that affect the spectral statistics are superpositions of such configurations. They are highly degenerate in energy, due to diluteness of the system and large number of possible configurations of photons in the extra sites, which results in peaks in DOS and deviations in the average ratio of consecutive energy gaps.
 
 For all photon numbers considered, the proportion saturates at some constant value as the size increases and the system becomes sufficiently dilute. Indeed, the saturation value is much larger for $N=3$ (approximately 70\%) than it is in cases of more photons (e.g. around 1\% for $N=8$). This explains why the special states have much stronger effects on the spectra of systems with small photon numbers. Such states are still present in large-$N$ systems, but their proportion is negligible and thus has practically no influence on the energy spectrum. We note that the proportion of special states depends on the ratio between the number of decorations and the number of sites $L_\mathrm{decor}/L_\mathrm{sites}$, which is equal to $1/2$ in the case shown here. More decorations would result in a larger proportion of these states.

%

\end{document}